\documentclass[twocolumn,showpacs,preprintnumbers,amsmath,amssymb,prb,floatfix]{revtex4}
\usepackage{graphicx}
\usepackage{dcolumn}
\usepackage{bm}

\begin{document}

\title{Critical Tunneling Currents in Quantum Hall Superfluids: \\ Pseudospin-Transfer Torque Theory}

\author{Jung-Jung Su$^{1,2}$}
\author{Allan H. MacDonald$^{1}$}
\affiliation{%
$^1$ Department of Physics, The University of Texas at Austin, Austin, TX 78712, USA}%
\affiliation{%
$^2$ Theoretical Division, Los Alamos National Laboratory, Los Alamos, New Mexico 87545,USA}%

\date{\today}

\begin{abstract}
At total filling factor $\nu=1$ quantum Hall bilayers can have an ordered ground state 
with spontaneous interlayer phase coherence.  The ordered state is signaled experimentally
by dramatically enhanced interlayer tunnel conductances at low bias voltages; at larger 
bias voltages inter-layer currents are similar to those of the disordered state.
We associate this change in behavior with the existence of a critical current
beyond which static inter-layer phase differences cannot be maintained, and examine the dependence of this critical current on 
sample geometry, phase stiffness, and the coherent tunneling energy density.  Our analysis is based in part on 
analogies between coherent bilayer behavior and spin-transfer torque physics in 
metallic ferromagnets.  Comparison with recent experiments suggests that
disorder can dramatically suppress critical currents.    
\end{abstract}

\maketitle

\section{Introduction}

At Landau level filling factor $\nu=1$ bilayer two-dimensional electron systems in the 
quantum Hall regime can have broken symmetry ground states\cite{fertig:1989,
macdonald:1990a,wen:1992,eisenstein:2004} with spontaneous inter-layer phase coherence.  These ordered states can be viewed 
as excitonic superfluids,\cite{macdonald:1990b,macdonald:2002} or as XY pseudospin ferromagnets\cite{moon:1995} 
in which the pseudospin is formed from the two-valued {\em which layer} quantum degree of freedom. 
The most robust experimental signature of these states, a vastly enhanced inter-layer 
tunnel conductance\cite{spielman:2000,jpeexpts,princeton,stuttgart} at small bias voltages, 
is still poorly understood from a quantitative point of view.

Two types of ideas, 
which differ most essentially in how the bias voltage is introduced in the theory, have been explored in an effort to 
understand the height and width of the tunnel conductance peak.  In one approach\cite{stern:2001,balents:2001,fogler:2001,brits}
the bias voltage is introduced as an effective magnetic field, uniform across the bilayer, which induces pseudospin precession 
around the $\hat z$ axis, driving the interlayer phase difference at a steady rate and inducing a purely oscillating 
inter-layer current.  When the microscopic inter-layer tunneling amplitude is treated as a 
perturbation, thermal and disorder fluctuations of the condensate are then responsible 
for
a finite {\em dc} conductance 
peak.  We refer to this type of theory below as the weak-coupling theory of the tunneling anomaly.
Weak-coupling theories predict\cite{stern:2001,balents:2001,fogler:2001,brits} splitting of the 
zero voltage tunnel conductance peak into separate finite voltage peaks in the presence of a 
magnetic field component parallel to the two-dimensional layers, an effect that is not\cite{spielman:2001} seen experimentally.
The second type of transport theory\cite{wen:1993,rossi:2004,khomeriki:2006,fil:2007,
su:2008} is formulated 
in terms of local chemical potentials of fermionic quasiparticles which may be altered by the ordered state condensate
but are still responsible for charge conduction.
In this type of theory the tunnel conductance is finite even in the absence of disorder and 
thermal fluctuations because charge has to be driven between normal-metal source and drain 
contacts.  The resistance generally   
depends\cite{rossi:2004,su:2008} on how the fermionic degrees of freedom which carry charge between leads
are influenced by order parameter and disorder configurations and cannot be 
described in terms of condensate dynamics alone.  In the second approach the 
width in voltage of the conductance peak is simply equal to the product of this 
resistance and the maximum current between source and drain at which the order parameter can
maintain a time-independent steady-state value.  

In this article we 
expand on the second type of theory of the conductance peak, using pseudospin-transfer 
torque ideas\cite{Stiles,slonczewski} borrowed from recent ferromagnetic-metal spintronics literature
to model the influence of the transport current on the pseudospin magnetization.  
We find that the critical current depends in general on details of the sample geometry and
on how disorder and localization physics influence transport inside the system.  Generally speaking 
however, the critical current is proportional to system area and to the inter-layer tunneling
amplitude when the condensate's Josephson length is larger than the system perimeter, and 
proportional to the system perimeter and to the square root of the inter-layer tunneling amplitude 
when it is shorter. 

Our paper is organized as follows.  In Section II we introduce the pseudospin-transfer torque 
theory of order-parameter dynamics in a bilayer quantum Hall ferromagnet.  We use this 
theory in Section III to discuss critical current values from a qualitative point of view.
In Section IV we report on a series of numerical studies
which take into account the two-dimensional nature of the systems of interest and 
the edge dominated current-paths typical of strong magnetic fields.  The
pseudospin transfer torque theory enables us to assess 
the influence of sample geometry on critical currents.
Finally in Section V we discuss the significance of our findings in relation to 
recent experiments.  We find that experimental critical currents 
are several orders of magnitude smaller than theoretical ones and argue that vortex-like
disorder-induced pseudospin textures must be largely responsible for this discrepancy.
We propose new experimental studies which can test our ideas and thereby 
achieve progress toward a quantitative theory of the spontaneous coherence tunneling anomaly.   

\section{Pseudospin Magnetism and the Landau-Lifshitz-Slonczewski Equation}

In the lattice model\cite{burkov} of $\nu=1$ bilayer systems, the local {\em which} layer degree of freedom can be expressed as a 
pseudospin defined by the operator
\begin{eqnarray}
\vec{S}_i = \frac{1}{2} \ \sum_{\sigma, \sigma'} 
a_{i,\sigma}^{\dagger} \vec{\tau}_{\sigma,\sigma'} a_{i,\sigma'}
\end{eqnarray}
where $i$ is the site index, $\sigma$ is the layer index and $\vec{\tau}$ is the Pauli matrix vector.
The pseudospin Hamiltonian has the form\cite{burkov} 
\begin{equation}
H_{\rm int} =\frac{1}{2} \sum_{i,j} (2 H_{i,j}-F^S_{i,j}) S_i^z S_j^z
-F^D_{i,j} (S_i^x S_j^x+S_i^y S_j^y)
\label{eq:englattice}
\end{equation}
where $H_{i,j} = \langle i,\sigma; j, \sigma' |V_{\rm col}| i,\sigma; j \sigma' \rangle$
is the direct Coulomb interaction associated with the $\hat{z}$ pseudospin 
component ({\em i.e.} with charge transfer between layers), 
$F^S_{i,j} = 
\langle i,\sigma; j, \sigma |V_{\rm col}| i,\sigma; j \sigma \rangle$,
is the exchange interaction between orbitals in the same layer and 
$F^D_{i,j} = 
\langle i,\sigma; j, {\bar \sigma} |V_{\rm col}| i,\sigma; j {\bar \sigma} \rangle$
is the exchange interaction between orbitals located in different layers.
Since $H_{i,j}$ is generally larger than $F^{S}_{i,j}$, the classical ground 
state is an easy-plane pseudospin ferromagnet with a hard $\hat{z}$ axis.  
In the limit of smooth textures the pseudospin energy functional 
has the form \cite{moon:1995} 
\begin{eqnarray}
\label{eq:engfunctional} 
E[\vec{m}]& =& \int d^2 r 
\left\{ \beta \, (m_z)^2 \right. \nonumber \\  && \left.
+ \frac{1}{2} \, \rho_s \, [ 
\left|\vec{\nabla} m_x\right|^2+ \left| \vec{\nabla} m_y\right|^2] 
-\frac{1}{2} \, \Delta_t \, n \, m_x 
\right\}
\end{eqnarray}
where $\vec{m}=\{m_x,m_y,m_z\}$ is the local pseudospin direction.
The parameters which appear in this expression are the anisotropy parameter $\beta > 0$, the pseudospin 
stiffness (or equivalently the exciton superfluid density) $\rho_{s}$, the splitting between symmetric and antisymmetric single-particle 
bilayer states due to interlayer tunneling  
$\Delta_t$, and the pseudospin density $n$.  

The mean-field-theory pseudospin ferromagnet\cite{fertig:1989}   
consists of a full Landau level of electrons in identical phase coherent bilayer states.
It follows that the mean-field-theory pseudospin density $n$ is equal to the full Landau level 
density, $(2 \pi l^2)^{-1}$.  (Here $l = (\hbar c/eB)^{1/2}$, where 
$B$ is the magnetic field strength, is the magnetic length.)   Mean-field-theory can also used\cite{moon:1995} to find explicit expressions for $\rho_s$ and $\beta$.  
In practice the values of these three parameters are modified\cite{joglekar:2001} by quantum and 
thermal fluctuations. 
The fourth parameter $\Delta_t$ is exponentially sensitive to the tunnel barrier between layers.
Parameters values can also be influenced by disorder on length scales shorter than those on 
which this coarse-grained continuum theory is applied; disorder on longer length scales would have
to be treated explicitly as we mention in the discussion section.  The upshot is that the numerical values of the parameters in Eq.(~\ref{eq:engfunctional})
are usually not accurately known and likely vary substantially from sample to sample.  
It is worth noting that $n$ must vanish at finite temperatures when $\Delta_t \to 0$
since two-dimensional systems cannot support spontaneous long-range phase order.  
Among all continuum model parameters the value of $\rho_s$, which is typically $\sim 10^{-4} {\rm eV}$,
is likely the most reliably known.        

Since the {\em which layer} pseudospin and the true electron spin have identical 
quantum mechanical descriptions, we can borrow from the ferromagnetic metal spintronics literature\cite{Stiles,slonczewski}  
and use the Landau-Lifshitz-Slonczewski(LLS) equation to describe how the semiclassical 
pseudospin dynamics is influenced by a transport current: 
\begin{eqnarray}
\label{eq:llscompact} 
\frac{d \vec{m}}{dt} =  
\vec{m} \times \vec{H}_{\rm eff} 
- \frac{ (\vec{j} \cdot \partial_{\vec{r}})\, \vec{m}}{n}-
\alpha \left( \vec{m} \times \frac{d \, \vec{m}}{d t} \right).
\end{eqnarray}
($\vec{j}$ is the number current density for electrons.) 
The second term on the right-hand-side of 
Eq.(~\ref{eq:llscompact}) 
captures the 
transport current effect.  Its role in these equations is similar to the role 
played by the leads in the pioneering analysis of coherent bilayer 
tunneling by Wen and Zee\cite{wen:1993} who argued that a term should be added to the 
global condensate equation of motion to account for the contribution of transport 
currents to the difference in population between top and bottom layers. 
Eq.(~\ref{eq:llscompact}) describes how a transport current alters the 
condensate equation of motion locally.
Its justification for the bilayer quantum 
Hall case is discussed in more detail below.
The essential validity of this expression has been verified by countless experiments. 
In the first term on the right hand side 
\begin{equation} 
\vec{H}_{\rm eff}=(2/\hbar n) (\delta E[\vec{m}]/\delta \vec{m})
\end{equation} 
describes precessional pseudospin dynamics in an effective magnetic field 
which is defined by the energy functional.  The third term 
accounts for damping of the collective motion due to coupling to 
its environment, in the present case the Fermi
sea of quasiparticles.  The damping term in 
Eq.(~\ref{eq:llscompact}) 
has the standard isotropic form used in the magnetism literature.  
A microscopic theory\cite{joglekar:2001a} of damping in quantum Hall bilayers
makes it clear that the damping is actually quite anisotropic.  We return to this point below. 

Magnetic order in the metallic ferromagnets to which this equation is normally applied 
is extremely robust, justifying the assumption that the spin-density magnitude is 
essentially unchanged even when the system is driven from equilibrium 
by a transport current.  The only relevant degree of freedom is the 
spin-density direction, whose dynamics is described by Eq.(~\ref{eq:llscompact}).  
We expect that the pseudospin transfer torque description of bilayer quantum
Hall systems will be most reliable when the order is most firmly established, 
that is far away from the phase boundary\cite{phaseboundary} that separates ordered and disordered states. 

The transport-current (Slonczewski\cite{slonczewski}) term in 
Eq.(~\ref{eq:llscompact}) 
can be understood in several different ways.  In the 
spintronics literature this term is normally motivated by an 
appeal to total spin conservation and referred to as the spin-transfer 
torque.  The idea is that when a ferromagnet's spin-polarized quasiparticles
carry a transport current through a spatial region with a non-collinear 
magnetization, they violate spin-conservation.  The collective 
magnetization must therefore compensate by rotating at a constant 
rate which is proportional to the fermion drift velocity.  
The form we use for the pseudospin transfer torque assumes that
each component of the pseudospin current is locally equal to the 
number current times the corresponding pseudospin direction cosine.
This property does not hold locally in a microscopic theory, but 
is expected to be valid in the smooth pseudospin texture limit we address.
(In metals an additional phenomenological factor is required to account 
for the difference in drift velocity between majority and minority spin electrons.)

Since pseudospin 
is not conserved in bilayer quantum Hall systems, as we can see
from Eq.(~\ref{eq:englattice}) or Eq.(~\ref{eq:engfunctional}), this 
argument does not apply directly.  If we appeal to a mean-field 
description of the pseudospin ferromagnet, however, we can obtain 
the same result by the following argument.  
Mean-field quasiparticles satisfy a single-particle 
Hamiltonian for a particle in a magnetic field which experiences both a scalar potential, and a 
pseudospin-dependent potential that can be interpreted as a
pseudospin effective magnetic field.  
Following standard textbook derivations it is possible\cite{currentcaveat} to derive an expression for the time-dependence of 
the contribution of a single quasiparticle to any
component of the pseudospin density.
The expression contains a pseudospin precession term and an additional term which is the divergence of the  
current of that component of pseudospin.  Summing over all quasiparticle states we obtain a precession term
that depends on the configuration of the order parameter, and hence on 
the pseudospin-field to which it gives rise, even in the 
absence of a current.  We obtain the 
additional current-driven term on the right-hand side of 
Eq.(~\ref{eq:llscompact}) 
when the pseudospin currents are non-zero and space dependent.  
The additional term in the equation of motion can also be 
viewed\cite{nunez} as a consequence of an altered relationship between 
pseudospin-dependent effective magnetic field and pseudospin polarization direction 
for quasiparticles that carry a current.  Eq.(~\ref{eq:llscompact}) should in principle also include a 
current related damping term\cite{duinestt} which we ignore in the present paper. 

When quantum Hall bilayer pseudospin ferromagnets are tilted far from their easy plane,
order tends\cite{champagne:2008} to be destroyed.  For that reason we are often most 
interested in the limit in which $m_z$ is much less than $1$.  It is therefore 
convenient to express the pseudospin direction in terms of the azimuthal angle $\phi$,
which is the inter-layer phase difference, and $m_z$ which is proportional to the 
layer polarization.  In terms of these variables the LLS equations take the form
\begin{eqnarray}\label{eq:lls}
\dot{m}_z &=&  
\left\{ -\frac{2}{n \hbar} \rho_s \, m_{\perp}^2 \, \vec{\nabla}^2 \phi 
+  \, \frac{\Delta_t}{\hbar} \, m_{\perp}  \sin\phi \right\}  \nonumber \\
&& - \left(\vec{v}_{ps} \cdot \vec{\nabla}) \, m_z \right\} 
+ \alpha_z \,m_{\perp}^2 \,\dot{\phi}
\nonumber \\
\dot{\phi} &=& m_z \, \left\{ 
\, \frac{2}{n \hbar} \, \rho_s  
\left( 
\left|\vec{\nabla} \phi \right|^2 
+ \frac{2}{m_{\perp}^4} \, \left|\vec{\nabla} m_z\right|^2 
+\frac{2m_z}{m_{\perp}^2} \, \vec{\nabla}^2 m_z \right) 
\right. \nonumber \\ && \left.
 - \frac{4}{n \hbar} \beta 
- \, \frac{\Delta_t}{\hbar} \, \frac{1}{m_\perp} 
\cos \phi \right\} 
\nonumber \\
&& - \left( \vec{v}_{ps} \cdot \vec{\nabla}) \, \phi \right\} 
- \frac{\alpha_\phi}{m_{\perp}^2} \,\dot{m}_z .
\end{eqnarray}
In Eq.(~\ref{eq:lls}) 
we have written $\vec{j}/n$, which has units of velocity, as the pseudospin velocity $\vec{v}_{ps}$. 

These equations do not on their own provide a closed description of pseudospin dynamics in the presence of electrical  
bias potentials and need to be supplemented by a theory which specifies the spatial dependence of the pseudospin  
current.  In general this quantity depends\cite{su:2008} on the order parameter configuration as well as on 
the contact geometry and 
external current or voltage biases.  The transport theory 
and the pseudospin dynamics theory are therefore not independent.  In the present paper we study 
voltage biased Hall bars with source and drain leads at opposite ends, 
and with a variety of shapes and sizes.  We assume, as a simplification, that the current distribution 
is defined by a local conductivity tensor with a large Hall angle.  
Given these simplifications, we are able 
to explicitly evaluate the maximum current at which time-independent order 
parameters are possible.  Because collective tunneling no longer contributes strongly to 
the {\em dc} interlayer current when the inter-layer phase is time-dependent, the 
interlayer conductance mechanism changes qualitatively when this maximum current is 
exceeded. We therefore associate this current with the experimental critical current.

\section{Approximate Critical Currents}

In this section we discuss approximate 
upper bounds on the critical current which are helpful 
in interpreting the numerical results described in the following section.
We use a simplified version of the static limit of the $\dot{m}_z$ LLS equation (Eq.(~\ref{eq:lls})) 
in which $m_z$ is assumed to be small:
\begin{eqnarray}
0=  -  \frac{\rho_s}{\hbar} \, \vec{\nabla}^2 \phi 
+ \frac{1}{2} \, \frac{\Delta_t n}{\hbar} \, \sin\phi - \frac{1}{2} \; \vec{j} \cdot \vec{\nabla} \, m_z .
\label{eq:steadystate} 
\end{eqnarray}
The three terms on the right-hand side can be identified as contributions to the time-dependence of  
$m_z$ (or equivalently of the exciton density) due respectively to the
divergence of the exciton supercurrent, coherent condensate tunneling, and the 
divergence of the $\hat{z}$ (layer antisymmetric or counterflow) 
fermionic pseudospin current.
The last contribution would be viewed as a spin-transfer torque in 
the analogous equations for an easy-plane anisotropy ferromagnetic metal. 

\begin{figure}[t]
\includegraphics[width=1\linewidth]{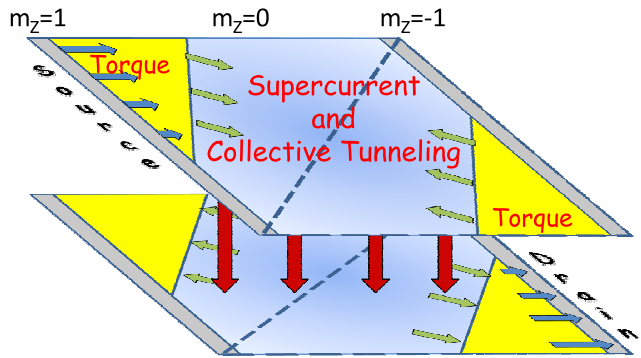}
\caption{(Color online) Separation of transport length scales in quantum Hall superfluid transport.
This theory is intended to apply when the ordered state is well established and Hall
angles are large because of the developing $\nu=1$ quantum Hall effect.
At large Hall angles current enters and leaves the samples at the 
{\em hot spot} corners, even when the source and drain contacts (gray) fully cover 
the ends of a Hall bar.  When order is well established, the
pseudospin orientation of the transport electrons achieves alignment with the condensate 
within a relatively small fraction of the sample area close to source and drain (solid yellow).  
In these areas pseudospin transfer torques convert  
transport currents into condensate counterflow supercurrents.
When source and drain are 
connected to opposite layers a net supercurrent is injected into the 
interior region of the sample. 
In the remaining sample area (shaded blue) collective interlayer tunneling 
can act as a sink for the counterflow supercurrent. 
}
\label{fig:mz}
\end{figure}
 
We start our qualitative discussion of critical currents by identifying some relevant 
length scales.  First the length scale,   
\begin{eqnarray}
\lambda = \sqrt{\frac{2 \rho_s }{\Delta_t n}},  
\end{eqnarray}
often referred to as the Josephson length because of the similarity between these 
equations and those which describe Josephson junctions, emerges from balancing the 
first and second terms.  In this paper we assume that the pseudospin 
magnetization direction departs from the $\hat{x}-\hat{y}$ plane only
over a small region close to the source and drain contacts whose spatial extent 
is small compared to the Josephson length.  (This issue is addressed again in the 
discussion section.)  If this is correct, we can separate length scales by 
identifying a region close to the contact which is small enough that we 
can ignore coherent tunneling by setting $\Delta_t n$ to $0$, and large enough that we can 
assume that $m_z$ is close to zero in the rest of the system.  (See Fig.~(\ref{fig:mz}).)   When $\Delta_t n \to 0$, Eq.(~\ref{eq:steadystate}) 
simply expresses the conservation of the sum of the excitonic and quasiparticle counterflow
currents.
Integration of Eq.(~\ref{eq:steadystate}) over the area close to the contact 
then simply describes conversion of quasiparticle counterflow current into 
condensate counterflow current.  The total counterflow current emerging from the 
area near the source contact is half of the total current flowing into the 
system since $m_z = \pm 1$ in the contact and $m_z \to 0$ far away from the 
contact.  In a tunnel geometry experiment the same counterflow current is 
generated near the source and drain contacts located at opposite dies of the sample, so the total counterflow
supercurrent injected {\em into} the system is equal to the total number current flowing {\em through} the system.

With this separation of length scales the quasiparticle 
(pseudospin transfer torque) term can be dropped in the remaining
portion $\tilde{A}$ of the total system area $A$. 
For static solutions of the LLS equations the condensate must satisfy an
elliptic sine-Gordon equation inside $\tilde{A}$:
\begin{eqnarray}
\lambda^2 \vec{\nabla}^2 \phi - \sin\phi=0.
\end{eqnarray}
When $\lambda \to \infty$ ($\Delta_t n  \to 0$), this equation states that
$\vec{\nabla}^2 \phi$ is zero.  It then follows from Green's theorem that no net
counterflow supercurrent can flow into the area $\tilde{A}$.  
In the tunnel geometry that means that time-independent 
order parameter values cannot be maintained in the presence of a transport
current unless $\Delta_t n \ne 0$.  The maximum tunneling current that can flow through the system is 
particularly simple to determine in the small $\Delta_t n$, large $\lambda$ 
limit.  When $\lambda$ is much larger than the system size
the angle $\phi$ cannot vary substantially over the 
system area.  With this simplification
the elliptic sine-Gordon equation can be integrated over the area $\tilde{A}$ 
to obtain 
\begin{equation} 
\label{smallcritcurrent} 
 \rho_s \int_{P} \vec{\nabla} \phi \cdot \hat{n} = \frac{\rho_s \tilde{A}}{\lambda^2}  \sin(\phi) 
\end{equation}
where $P$ is the perimeter of $\tilde{A}$ and $\hat{n}$ is proportional to the outward 
normal.   The left hand side of Eq.(~\ref{smallcritcurrent}) is the 
net supercurrent which flows out of the region $\tilde{A}$ from its boundaries near the 
source and drain contacts, identified above as the number current flowing through
the system.  Since the maximum value of $|\sin(\phi)|$ is $1$, it follows that 
the maximum current consistent with a time-independent order parameter in this case is
\begin{equation} 
I^{c}_{B} = \frac{e \tilde{A} \rho_s}{\hbar \lambda^2} = \frac{e \tilde{A} \Delta_t n}{2 \hbar}.
\end{equation}  
Since the critical current in the small $\Delta_t n$ limit is proportional to the 
area of the system we will refer to this quantity as the bulk critical current, as 
suggested by the notation used above.  

For larger $\Delta_t n$, $\lambda$ is no longer larger than the 
system size and it is not possible to maintain the maximum value of 
$\sin(\phi)$ across the system.  The LLS equation critical current in this regime 
depends on geometric details and we have not been able to obtain rigorous bounds.
We can make a rough estimate by 
following an argument along the following lines.  The elliptic sine-Gordon
equation is very similar to the regular sine-Gordon equation in which 
second order derivatives with respect to time and position appear with 
opposite signs.  This 1+1 dimensional sine-Gordon equation appears as the 
Euler-Lagrange equation of motion of a system with a Lagrangian with a 
kinetic-energy term proportional to $\rho_s (\partial_t \phi(x,t))^2$ and a 
potential energy term proportional $\Delta_t n \cos(\phi(x,t)$.  
Since total energy (integrated over position $x$) is conserved by this 
dynamics it follows that the variation of the typical value of 
$\rho_s (\partial_t \phi)^2$ along the space-time boundary cannot be 
larger than $\sim \Delta_t n$.  When this energy conservation condition
is mapped from the regular sine-Gordon equation to the (imaginary time) 
elliptic sine-Gordon equation we can conclude that the typical
value of $\vec{\nabla} \phi \cdot \hat{n}$ along the boundary of $\tilde{A}$ 
near the source contact cannot differ from the typical value of $\vec{\nabla} \phi \cdot \hat{n}$
along the boundary near the drain contact by more than $\sim \Delta_t n$.
It follows that the current flowing through the system from source to drain should not 
be much larger than  
\begin{equation}
I_{E}^{c} \sim \frac{e W \rho_s} {\hbar \lambda} 
\end{equation} 
where $W$ is the length of the 
contract 
region, or the width of a Hall bar assumed to be 
contacted at its edges.  We will refer to $I^{c}_{E}$ as the edge critical current,
since it is limited by the length of one edge.
For stronger interlayer coupling then 
critical current is expected to vary as $(\Delta_t n)^{1/2}$ once $\lambda$ is smaller than the 
Hall bar length.  Finally we note that because of 
hot-spot effects in transport with large Hall angles the pseudospin transfer torque 
will act at the sample corners.  Since the supercurrent is converted into coherent pseudospin precession over the length 
scale $\lambda$, the effective size of the contact region will be $\sim \lambda$ when the Hall angle is large and 
$\lambda$ is smaller than $W$.
We therefore estimate that the critical current is close to 
\begin{equation}
I^{c}_{C} \sim \frac{e \rho_s}{\hbar}, 
\end{equation} 
independent of $\Delta_t n$, under these circumstances.  
We refer to this last critical current as the 
corner critical current.  In the following section we compare numerical LLS 
critical currents with these rough estimates.

\section{Model Calculations}

The numerical calculations we describe below are similar to those 
carried out in micromagnetic descriptions of ferromagnetic metal spin-transfer torque physics, but 
are applied here to pseudospin transfer physics in condensed bilayers.
We divide the system area into pixels within which the pseudospin magnetization
is assumed to be constant.  Except where noted we used square $10 l \times 10 l$ pixels
where $l$ is the magnetic length. 
We rewrite the spin-transfer torque term in Eq.(~\ref{eq:lls}) in the discretized form:
\begin{eqnarray}
\left.\dot{m}_{\alpha}\right|_{ST} &=&
\sum_k \frac{ I_n }{n A_{\rm pixel}} 
\ [ m_{k,\alpha}-(\vec{m}_k \cdot \vec{m}) m_{\alpha}]  
\end{eqnarray}
where $k$ labels the four neighboring sites, $A$ is the pixel area,
and $\alpha$ labels components of the magnetization orientation in the pixel of interest.
$I_n$ is the quasiparticle number current flow from neighbor site $n$ into this pixel.
In this article we estimate values of $I_n$ by solving the resistor network model
obtained by discretizing a continuum model with a local conductivity that includes a Hall component.
\begin{eqnarray}
\{I\} = [G]\{V\}
\end{eqnarray} 
where $[G]$ is the conductance matrix that describes the local conductivity 
including Hall components, $\{V\}$ is a vector of local voltages in each pixel, and $\{I\}$ is a vector composed of currents
that flow between pixels.  
Using current conservation conditions the dimension of the matrix can be reduced to the pixel number $N$. 
Given the source and drain contact voltages, we can  
solve for the internal voltage and current distributions and for the variation of source and 
drain currents across the contacts.  Note that quantum Hall physics comes into play through the Hall 
contributions to the conductivity matrix $[G]$. 
The local Hall conductivity was set to $\sigma_{xy}=e^2/h$, which is close to the appropriate value for $\nu=1$ whether 
or not the Hall plateau is fully formed, and the longitudinal conductivity was set to 
\begin{equation}
\sigma_{xx} = 0.05 \, {\rm exp}(-m_z^2/W)\, (\vec{m} \cdot \vec{m}_L)  (\vec{ m} \cdot \vec{ m}_R) \, e^2/h
\end{equation} 
where $\vec{m}_{L,R}$
is the magnetization orientation to the left and right of a boundary separating two-pixels. 
The Hall angle used in these 
calculations was therefore $\tan^{-1}(20)$ over 
the largest part of the sample in which the pseudospin magnetization 
is close to collinear and planar.  The results we report on are not sensitive to the 
Hall angle, provided that it is large.    

The current which flows into a pixel is 
assumed to have the same pseudospin polarization as the pixel from which it 
is incident.  Currents entering or exiting from the contacts are 
assumed to have $m_z=1$ for top layer contacts and 
$m_z=-1$ for bottom layer contacts. 
In this way the pseudospin transfer torque and the Landau-Lifshitz precessional
torque acting on each pixel's pseudospin can be evaluated.  
Note that the pseudospin torque depends not only on current paths, but 
also on the 
pseudospin magnetization configuration.

\subsection{Critical Current Identification}

As mentioned in the previous section, we identify the critical current  
as the circuit current value above which a time independent solution no longer exists. 
When the external current is small, the Gilbert damping term in the LLS equation
relaxes the pseudospin magnetization into time-independent configurations.
The behavior of the pseudospin as the current increases is partly analogous to the 
behavior of a damped pendulum driven by an increasingly strong torque.

We identify the critical current numerically by slowly increasing the 
driving voltage (by $\delta V$ per time step) and monitoring the change
in magnetization. 
To be more explicit, we examine 
\begin{equation}
\label{eq:deltam}
\delta m \equiv \sum_{i}
|\vec{m}_{i} (V+\delta V)-\vec{m}_{i} (V)|/N
\end{equation} 
\begin{figure}[t]
\includegraphics[width=1.\linewidth]{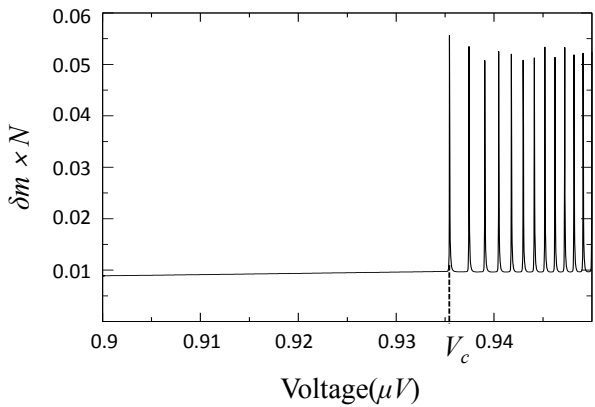}
\caption{Magnetization change per time step 
$\delta m \times N$ {\em vs.} applied voltage in a $500 l \times 400 l $ system 
for $2\pi l^2 n \Delta_t = 10^{-5} E_0$.  This curve was obtained by changing the bias voltage 
by $\delta V=2.5 \times 10^{-5}$ $\mu$V at each time step.
$E_0=e^2/\epsilon l$, the energy unit used in all our calculations has a   
typical experimental value $\sim 7$ meV.  
$\delta m$ develops large amplitude oscillation when the voltage exceeds the critical voltage $V_c$.
}
\label{fig:typicalIc}
\end{figure}
The sum in Eq.(~\ref{eq:deltam}) is over all pixels.
The critical current can be defined precisely as the current above which $\delta m$ remains finite when 
$\delta V \to 0$.  In practice we choose a suitably small value of $\delta V$ and 
examine the current or voltage dependence of $\delta m$. 
We find that $\delta m$ increases dramatically and begins oscillating at a voltage 
we identify as the critical voltage.  (See Fig.(~\ref{fig:typicalIc}).)
An alternative but more laborious method of obtaining critical currents 
is to sequentially examine the dynamics of the pseudospin magnetization at a 
series of fixed values of the applied voltage $V$. 
If the applied voltage is below its critical value, $\delta m$ will approach zero exponentially at large 
times.  If the applied voltage is above its critical value $\delta m$ will not approach zero and 
usually exhibits an oscillatory time dependence. 
In our calculations we used the first approach to determine an approximate value of the 
critical voltage (and hence the critical current) and the second method to refine 
accuracy.  


\subsection{$I_c$ {\em vs.} $\Delta_t$}

It is useful to start by briefly discussing the single-pixel limit of the calculation, which should apply approximately to the case in which 
the Josephson length is longer than the system size.  The steady limit of the  
LLS equation for the $\hat{z}$-component of the pseudospin is  
\begin{eqnarray}
0=\frac{1}{2} \frac{\Delta_t n A_{\rm pixel}}{\hbar} \sin\phi
-\frac{I}{2e} \ (m_{z,L}-m_{z,R})
\end{eqnarray} 
where $I$ is the charge current flowing through the system and 
$m_{z,L}$ and $m_{z,R}$ are the $z$ component of pseudospin for the source and drain leads at the left and right ends of the 
sample. For the drag geometry (current entering and exiting from the 
same layer) $m_{z,L}=m_{z,R}$, there is no spin torque term in the 
single-pixel calculation, and the steady state equation can be satisfied by setting $\sin\phi=0$. 
For the tunneling geometry $m_{z,L}=-m_{z,R}$, the maximum pseudospin torque that can be
compensated by the tunneling term is obtained by setting $\sin\phi \to 1$.  We therefore obtain 
\begin{eqnarray}
I^{c} = I^{c}_{B} = \frac{e}{2 \hbar} \, \Delta_t n A_{\rm pixel}
\end{eqnarray}
This gives a linear dependence of $I_c$ on the single-particle tunneling strength $\Delta_t$. 
The numerical procedures described above accurately reproduce this simple result.

\begin{figure}[t]
\includegraphics[width=1.\linewidth]{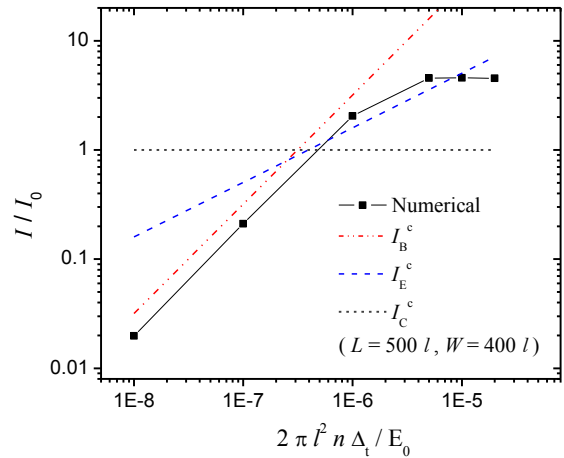}
\caption{(Color online) Critical current {\em vs.} $2\pi\ell^2 n \Delta_t$ in a $400 l \times 500 \ l$ system. 
The dash-dot (red), long-dash (blue) and short-dash (black) curves plot the values of the bulk
($I^{c}_{B}$), edge ($I^{c}_{E}$) and corner ($I^{c}_{C}$) limited critical currents discussed in the 
text for this sample geometry.  The square dots plot LLS equation critical currents at a 
series of $\Delta_t n$ values.  All the calculations in this paper were performed using 
pseudospin stiffness (exciton superfluid density) $\rho_s = 0.005 E_0$ where $E_0=e^2/\epsilon \, l$ is 
the energy unit.  
Currents are in units of $I_0 = I^{c}_{E} = e \rho_s /\hbar $.  For the value of 
$\rho_s$ used in these calculations $I_0 \simeq  8 n$A.}
\label{fig:IcvsDelta(2D)}
\end{figure}


As explained previously and discussed more fully later, we believe that the pseudospin transfer torque in most quantum Hall superfluid
experiments acts in a small fraction of the system area.
We therefore need to perform calculations with many pixels in order to represent a typical measurement.
Fig.~(\ref{fig:IcvsDelta(2D)}) shows numerical critical current results for a 
fixed sample geometry as a function of $\Delta_t$.  In this figure $I_0= I^{c}_{C} = e \rho_s /\hbar \sim 8 n$A is the 
unit of current and $E_0=e^2/\epsilon \, l$ is the unit of energy. 
Typical quantum Hall superfluid experiments are performed at a magnetic field of roughly 2.1 Tesla for 
which the magnetic length is 17.65 nm which gives $E_0=6.4$ meV.
In all the calculations reported on here we used a    
pixel area $A_{\rm pixel} = 10 \times 10 \, l^2$ and the 
the mean-field theory estimate\cite{moon:1995} ($\rho_s \simeq \, 0.005 E_0$) for the pseudospin stiffness. 
The calculations in Fig.(~\ref{fig:IcvsDelta(2D)}) are for 40 pixels in the width $W$ direction and 50 pixels in the 
current $L$ direction.   For this system size , the Josephson length is approximately equal to $L$  
when $2 \pi l^2 \Delta_t n \sim 4 \times 10^{-7} \, E_0$. 
The numerical results illustrated in Fig.~(\ref{fig:IcvsDelta(2D)}) show the  
crossovers from $\Delta_t$-dependence, to $\Delta_t^{1/2}$-dependence, to saturation 
as $\Delta_t$ increases that was anticipated in our qualitative discussion.  
At small $\Delta_t n$ the critical current is reduced by a small fraction compared
to the single pixel result in accord with the Fig.(1).

\begin{figure}[t]
\includegraphics[width=1.\linewidth]{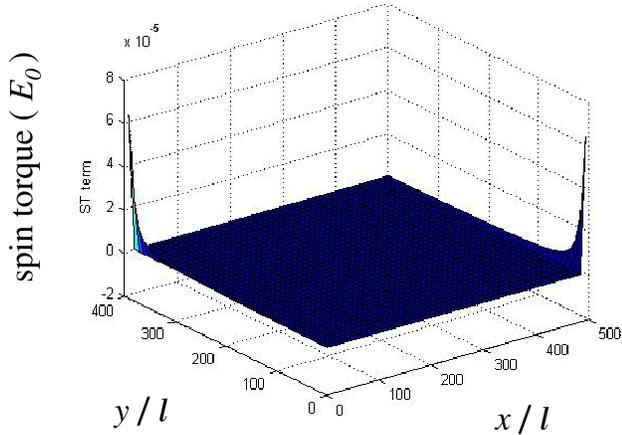}
\caption{Spatial distribution of the $\hat{z}$ component of the 
pseudospin transfer torque in a system with $500 \times 400 \ l^2$ area, $2\pi l^2 n \Delta_t = 10^{-6} E_0$,
and $I/I^{c} = 0.29$.
The pseudospin transfer torque in the model studied here acts mainly in the hot spot pixels. 
$E_0$ and $l$ are defined as in previous figures.
}
\label{fig:mzEOM_ST}
\end{figure}

\begin{figure}[t]
\includegraphics[width=1.\linewidth]{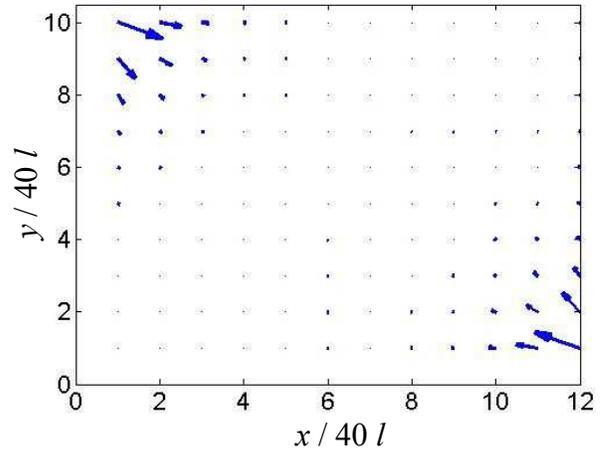}
\caption{Supercurrent distribution in system with area $500 \times 400 \ l^2$, $2\pi l^2 n \Delta_t = 10^{-6} E_0$,
and $I/I^{c} = 0.29$.
This plot is for a tunneling geometry in which the source is a top layer contact and the drain is a bottom layer 
contact.  Supercurrents are generated near both hot spots and flow diagonally toward the 
sample center. 
$E_0$ and $l$ are defined as in previous figures. 
}
\label{fig:Tnnl_SC}
\end{figure}   

\begin{figure}[t]
\includegraphics[width=1.\linewidth]{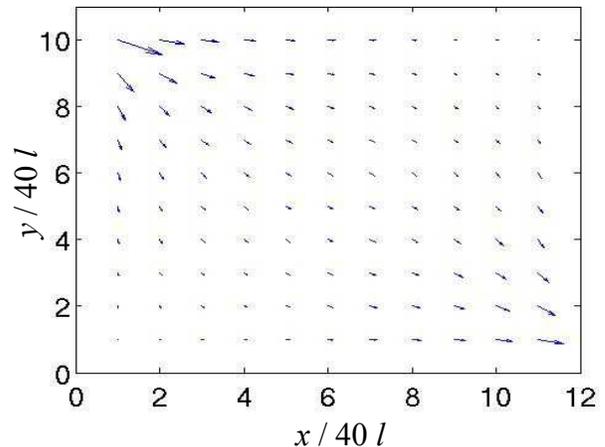}
\caption{Supercurrent distribution in system with area $500 \times 400 \ l^2$, $2\pi l^2 n \Delta_t = 10^{-6} E_0$.  
This plot is for the drag geometry case in which the source and drain are both 
top layer contacts, but other parameters of the calculation are identical to those used
in the preceding tunnel-geometry figure. $E_0$ and $l$ are defined as in previous figures.  
}
\label{fig:Drag_SC}
\end{figure}  

We now examine the ingredients which enter these numerical results in greater detail.
According to the schematic Fig.~(\ref{fig:mz}), the pseudospin transfer torque acts only
near the hot spots at which current enters and exits the sample.  
Fig.~(\ref{fig:mzEOM_ST}) shows a typical numerical results for the spatial distribution of the pseudospin transfer 
torque.  In the present model, the area of the region in which 
transport current is converted into supercurrent depends on the pixel size and 
the pseudospin stiffness and anisotropy coefficients.  In the tunneling geometry, 
counterflow supercurrent is generated near both source and drain hot spots and flows diagonally
toward the center of the sample.  In Fig.~(\ref{fig:Tnnl_SC}) we show a typical supercurrent distribution 
for the tunneling geometry case.  The corresponding distribution for an equivalent drag experiment 
(both contacts connected to the same layer) 
is illustrated in Fig.~(\ref{fig:Drag_SC}). 
Since the elliptic sine-Gordon equation applies locally when
the pseudospin transfer torque is negligible, it follows from 
Green's theorem and Eq.(~\ref{eq:steadystate}) that in a steady state the total counterflow current injected into
the interior of the sample (which for the tunnel geometry equals the total charge current flowing 
through the system) must match the area integration of $(1/2) (\Delta_t n /\hbar) \sin \phi$. 
For a drag geometry experiment the same integral should vanish.

\begin{figure}[t]
\includegraphics[width=1.\linewidth]{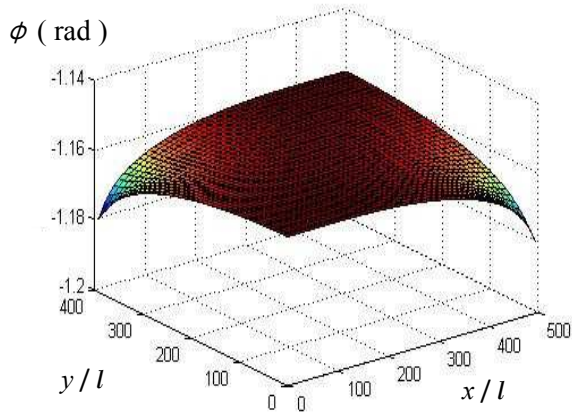}
\caption{Pseudospin phase distribution in a system with area $500 \times 400 \ l^2$, $2\pi l^2 n \Delta_t = 10^{-8} E_0$ and 
$I/I^{c}=0.75$.  The Josephson length at this value of $\Delta_t$ 
is $\sim 2500 l$. Note that $\phi$ is roughly constant through the system and that its value is close to  
$\pi/2$ because $I$ is close to $I_c$.  The units used here are the same as in previous figures.  
}
\label{fig:phi_smallDelta}
\end{figure}   

\begin{figure}[t]
\includegraphics[width=1.\linewidth]{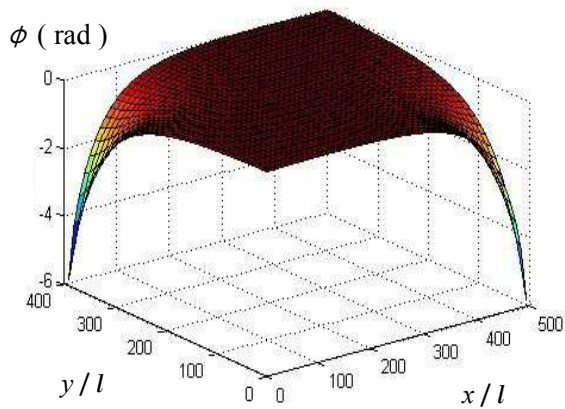}
\caption{Phase distribution of pseudospin in a system with area $500 \times 400 \ l^2$, $2\pi l^2 n \Delta_t = 1.2\times 10^{-5} E_0$
and $I/I^{c} = 0.97$ . The Josephson length at this value of $\Delta_t$ is $\sim 70 l$. 
Note that a steady state is reached even though $\phi$ varies considerably across the sample and has values larger than $\pi/2$.
The units used here are the same as in previous figures.}
\label{fig:phi_largeDelta}
\end{figure}   

We now examine steady state condensate configurations at currents near the critical current for
both small $n \Delta_t$ and large $n \Delta_t$ limits.  In the former case our expectation that
$\sin \phi$ should be nearly constant except near the hot spots is confirmed in 
Fig.~(\ref{fig:phi_smallDelta}).  Collective tunneling acts sinks the injected counterflow 
supercurrent at a rate that is nearly constant across the sample area.  The large $n \Delta_t$ case is more complex.
$\phi$ varies over a large range and $\sin(\phi)$ changes sign in different parts of the sample.
Near the sample corners $\phi$ varies rapidly with position because of the 
large counterflow supercurrents.  In the interior of the sample the phase changes 
less rapidly because the flow pattern spreads and because 
coherent tunneling is providing the required current sink.
The critical current in the large $n \Delta_t$ limit, 
depends in a complex way on the geometry of the sample and on the spatial distribution
of the pseudospin transfer torques.  
Nevertheless, the critical current saturation we find in our 
numerical studies suggests that once the Josephson length is smaller compared to both the 
width and the length of a Hall bar sample with a large Hall angle,
it is no longer relevant to the critical current value. 


\subsection{$I_c$ {\em vs.} System Geometry}

\begin{figure}[t]
\includegraphics[width=1.\linewidth]{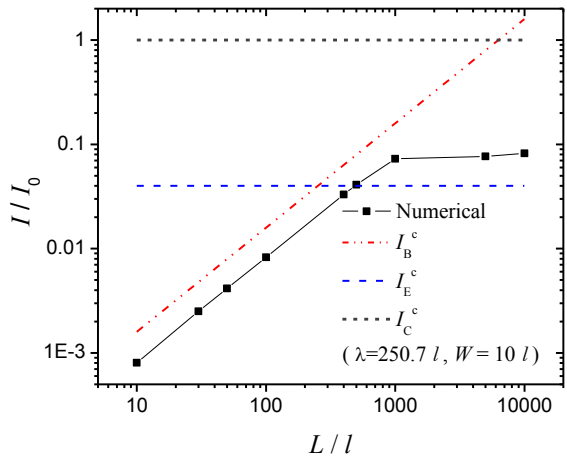}
\caption{Critical current {\em vs.} system length $L$ for a narrow Hall bar with width $W=10 \ l$. 
The single-particle tunneling amplitude $2\pi l^2 n \Delta_t = 10^{-6} (E_0)$ corresponding to $\lambda \sim 250 l$ }
\label{fig:IcvsL(narrow)}
\end{figure}

\begin{figure}[t]
\includegraphics[width=1.\linewidth]{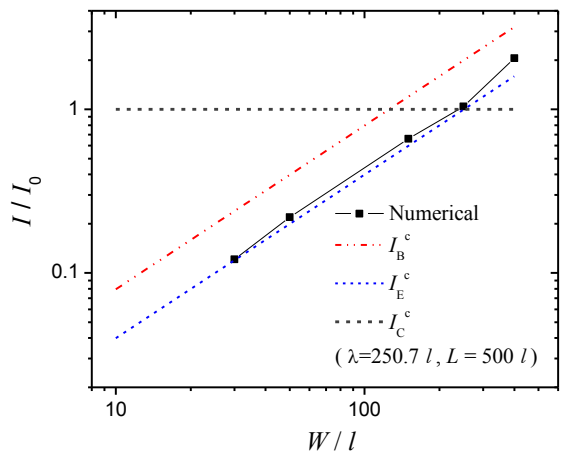}
\caption{Critical current v.s. system width $W$ for length $L=500 \ l$). The single-particle tunneling amplitude 
$2\pi l^2 n \Delta_t = 10^{-6} (E_0)$, corresponding to $\lambda \sim 250 l$ }
\label{fig:IcvsW}
\end{figure}

Finally, we briefly discuss critical current dependence on Hall bar dimensions at 
fixed $n \Delta_t$.  
In Fig.(~\ref{fig:IcvsL(narrow)}) we plot $I^c$ {\em vs.} Hall bar length in a series of model samples with a 
fixed single pixel width $W= 10 \, l$ and single-particle tunneling amplitude $2 \pi l^2 n \Delta_t= 10^{-6} E_0$.
For these parameters the $W$ is much smaller than the Josephson length. 
The critical current increases linearly with sample length and is therefore proportional to sample area 
until it saturates at $L \sim 1000 \, l$.  The length at which the current saturates is a bit longer
than the Josephson length and the value of the critical current is accordingly somewhat larger than the 1D 
estimate\cite{su:2008} $I^{c}_{E}$.  The large $L$ behavior is however consistent with the 
expectation that the critical current should not increase with system length once 
$L$ is substantially larger than the Josephson length $\lambda$. 
In Fig.~(\ref{fig:IcvsW}) we plot $I_c$ {\em vs.} system width $W$ with the length fixed at $500 \, l$
and the same $\Delta_t$ as above.  The range of $W$ covered is limited somewhat by 
numerical practicalities and goes from a width much smaller than
the Josephson length to a width which is somewhat larger.  Over this range deviations from
the 1D model in which the critical current is proportional to Hall bar width
are small. 
 
\section{Discussion and Conclusions}

We start our discussion by commenting briefly on some essential 
differences between the critical current of a Josephson junction and 
critical currents for coherent bilayer tunneling.
In a Josephson junction, current can flow without dissipation across a 
thin insulating layer that separates two superconductors.
The difference in condensate phase 
across the junction $\phi_{J}$ is normally zero in 
equilibrium but can be driven to a non-zero 
steady state value when biased by current flow $I_{J}$ in the circuit 
in which the junction is placed.  For thick insulating layers the  
current is related to the phase difference by 
\begin{equation} 
\label{eq:jj} 
I_{J} = I^{c}_{J} \; \sin(\phi_{J}),
\end{equation} 
where $I^{c}_{J}$ is the junction's critical current.  
Eq.(~\ref{eq:jj}) should be compared with Eq.(~\ref{eq:steadystate}).  
The most obvious difference is the appearance of the 
lateral 2D coordinate in the coherent bilayer case. 
In the Josephson junction case, the lateral dependence of $\phi_{J}$
usually plays no role unless an external magnetic field is present.
In the coherent bilayer case, on the other hand, lateral translational
invariance is always broken because the pseudospin transfer torques 
that ultimately drive the coherent tunneling current do not 
act uniformly across the system.

A closer comparison is 
possible in the special case in which the pseudospin stiffness is large enough 
to inhibit lateral variation of $\phi$.  Integration of Eq.(~\ref{eq:steadystate})
over the area then yields for the coherent bilayer 
\begin{equation} 
\label{eq:bilayerjj}
I_{BL} = I^{c}_{B} \; \sin(\phi)  
\end{equation} 
where $I^{c}_{B}$ is the bulk critical current discussed in the body of the paper and 
\begin{equation}
\label{eq:Iin} 
I_{BL}= e \, \int_{P} \big[ \frac{\rho_s}{\hbar} \vec{\nabla} \phi  + \frac{m_z}{2} \; \vec{j} \, m_z \big] \, \cdot \hat{n}. 
\end{equation} 
is the injected counterflow current.  The most essential difference between tunneling in 
coherent quantum Hall bilayers and Josephson junctions lies in the difference in physical 
content between the bias currents $I_{J}$ and $I_{BL}$.  In the case of a Josephson junction it is a 
bulk dissipationless 
supercurrent which flows perpendicular to the plane of the junction.  In the case of a coherent quantum Hall 
bilayer, it is the counterflow ($\hat{z}$) component of the quasiparticle current.
The counterflow component of the quasiparticle current is 
normally fully converted to condensate counterflow supercurrents by 
pseudospin transfer torques, as we have discussed at length.  
The voltage drop across a Josephson junction can be measured and vanishes
below the critical current.  Because fermionic quasiparticle 
transport, transverse counterflow supercurrents, and 
collective interlayer tunneling are unavoidably 
intertwined in the coherent bilayer tunneling case, there is no 
corresponding measurable voltage which vanishes below the
critical current.  Instead, the critical current is marked experimentally
by an abrupt increase in measured resistances.

Next we draw attention to the significance of the qualitative difference
in experiment between drag geometry and tunnel geometry transport measurements.  
In our theoretical picture is correct there should be no qualitative difference
difference between voltage measurements in these two geometries at currents below the critical current, 
once coherence is well established.  
(Important differences between these measuring geometries 
can occur in the interesting regime close to the phase boundary where 
interlayer coherence may be established over a small fraction of the sample area.\cite{sternhalperin}) 
Pseudospin transfer torques in tunnel and drag geometry experiments 
with similar total circuit currents give rise to similar counterflow supercurrents,
as seen in Figs. (5) and (6). 
Excitonic condensates have a maximum local (critical) supercurrent density 
that they can support, which in the case of 
quantum Hall bilayer exciton condensates has been estimated\cite{radzihovsky} 
to be $\sim 1 {\rm A} {\rm m}^{-1}$.  For a typical sample width of 
$10^{-4}$m this corresponds to a critical current in the ${\rm mA}$ regime, 
orders of magnitude larger than the current levels typically employed in quantum Hall experiments.
Even accounting for possible corrections due to disorder, 
local critical currents are unlikely to be approached experimentally
for either contact geometry. (If they were, we would expect more 
similarity between the two experiments.)  The critical currents 
which are important for typical tunneling experiments are not local critical currents,
but global critical currents which set an upper limit on the rate at 
which injected supercurrent can be sunk by collective tunneling.
Since there is no net injected supercurrent in the drag geometry 
experiment, only local limits apply.  In the tunnel-geometry 
experiment, the global limit must be satisfied, setting a 
critical current scale which can be orders of magnitude smaller if 
$\Delta_t$ is small.  Although the relationship anticipated here between 
tunneling geometry and drag geometry transport has not been specifically tested 
experimentally, it appears to us that it is clearly consistent with published data.

We now turn to a comparison of our theory with experiment, and in particular with 
the recent experiments of Tiemann {\em et al.}\cite{Tiemann:2009}
who have systematically studied the dependence of the tunneling critical currents in 
their samples on temperature and layer separation.
The data of Tiemann {\em et al.} appear to be broadly consistent with 
that reported in earlier \cite{spielman:2000,spielman:2001} but focus more 
on tunneling anomalies in the region of the phase diagram far
from the coherent state phase boundary.  Tiemann {\em et al.} find i) that 
the critical tunneling currents in their samples are 
proportional to system area, ii) that their typical value is
$\sim 10 {\rm nA}\,{\rm mm}^{-2}$, and that they saturate 
upon moving away from the coherent state phase boundary\cite{phaseboundary}  
either by lowering temperature or by decreasing the ratio of layer 
separation to magnetic length.  

Tiemann {\em et al.}'s finding that the 
critical current is proportional to area would be consistent with our analysis 
if the Josephson length was comparable to or longer 
than the $\sim 10 {\rm mm}$ length of their long-thin Hall bars.
When in the bulk-limited critical current regime, the critical current 
should be given by $I^{c}_{B}$
\begin{equation} 
\frac{I^{c}_{B}}{A} = \frac{e \tilde{\Delta}_t}{4 \pi l^2 \hbar} = 3.1 B [{\rm Tesla}] \times \tilde{\Delta}_t [{\rm eV}] \times 10^{4} {\rm A}\, {\rm mm}^{-2}
\end{equation} 
where $\tilde{\Delta}_t = 2\pi l^2 \Delta_t n$ is the interlayer tunneling amplitude 
suitably renormalized by quantum and thermal fluctuations and $B$ is the magnetic
field strength at $\nu=1$.  Inserting the magnetic field strength, experimental 
critical currents can be recovered by setting $\tilde{\Delta}_t \to 10^{-13} {\rm eV}$.
Therein lies the rub.  Although values for $\tilde{\Delta}_t$ on this scale 
do imply Josephson lengths that are on the ${\rm mm}$ length scale and therefore 
consistent with bulk-limited critical currents in the samples 
studied by Tiemann {\it et al.}, they are four or more orders of magnitude 
smaller than values expected on the basis of theoretical estimates.  
Suspicion that there is a fundamental discrepancy is established more 
convincingly, perhaps, by observing that these critical currents also
correspond to $\tilde{\Delta}_t$ values several orders of magnitude  
smaller than seemingly reliable experimental estimates\cite{spielman:thesis} 
based on the inter-layer tunneling conductance in similar samples at zero-field.  
The key issue then in comparing critical current theories with experiment 
appears to be explaining why they are so small.  
 
We expect on general grounds that the experimental value of 
$\Delta_t n \equiv \tilde{\Delta}_t/2\pi l^2$ should be renormalized downward by both 
thermal and quantum fluctuations and by disorder.  Indeed according to the 
familiar Mermin-Wagner theorem, the 
order parameter $n$ must vanish for $\Delta_t \to 0$ at finite temperatures.   
The importance of thermal fluctuations is strongly influenced by 
$k_B T/\rho_s$.  If mean-field theory 
estimates\cite{moon:1995} 
of $\rho_s$
can be trusted, the value of $\rho_s$ in typical experimental samples should be 
$\sim 3 \times 10^{-5} {\rm eV}$ and $k_B T/\rho_s$ should therefore be 
less than $0.1$ at the lowest measurement temperatures.  At these 
low temperatures thermal fluctuations alone appear to be
insufficient\cite{pesin} to explain the discrepancy even
though $\Delta_t/k_BT$ is certainly small.  The experimental finding 
that the critical current saturates at low temperatures supports this 
conclusion.  Similarly, theoretical estimates that quantum fluctuation corrections to the 
order parameter are not\cite{joglekar:2001} large 
well away from the transition boundary are consistent with the 
experimental finding of saturating critical currents in this 
regime.  It appears that an explanation for the small critical currents 
must be found in the disorder physics of quantum Hall superfluids.

The analysis of tunneling critical currents presented in this paper   
can accommodate disorder implicitly through its 
influence on the parameters $\rho_s$ and $\Delta_t n$.  
Including disorder effects through renormalized coupling 
constants would be adequate if the characteristic length scales for 
disorder physics are smaller than characteristic length 
scales like $\lambda$ which are relevant for pseudospin transfer 
torques.  In quantum Hall superfluids disorder may play a more 
essential role by nucleating charged merons\cite{meronrefs} (vortices). 
As we have explained, provided that the pseudospin transfer torque
acts only close to the source and drain, the critical current is 
proportional to the integral of $\Delta_t n$ over the area of the sample.
This integral is reduced to zero by a single undistorted meron located at the center 
of a large sample.  It has in fact often\cite{stern:2001,balents:2001,fogler:2001,eastham:2009,fertig:2005,roostaie:2008}
been recognized that disorder induced vortices might play 
an essential role in many of the transport anomalies associated with bilayer 
coherence.  It seems likely to us that this type of physics is very likely responsible for the 
small critical currents seen in experiment, but that existing theory is unable to 
account for this effect quantitatively.  The current status of the subject calls for 
a detailed analysis of how they influence critical currents.  On the experimental side, the importance of complex
disorder-related pseudospin textures for critical current values could be reduced and 
the essential physics which limits critical currents revealed more clearly by 
studying samples with much larger bare values of $\Delta_t$.

Although we have attributed the substantial quantitative disagreement between 
pseudospin transfer torque theory and experiment to disorder-induced 
pseudospin textures and have suggested a strategy for achieving more quantitative 
tests of our theory, it is appropriate to step-back and reconsider 
other theoretical pictures that might be relevant to coherent bilayer tunneling 
experiments under some circumstances.  For example, the version of the 
pseudospin-transfer torque theory applied here is based on the 
simplest possible assumption for the local pseudospin-polarization of the 
transport current, namely that the pseudospin current polarization simply follows
the pseudospin density polarization.  This assumption is certainly not generically 
correct, but its replacement requires more detailed knowledge of fermionic 
quasiparticle transport behavior.  One approach is to assume that the 
quantum Hall effect establishes edge state transport and use experimental 
voltage probe data to infer\cite{yoshioka} the length scale on which the 
pseudospin polarization of injected electrons is relaxed, and therefore the 
length scale over which the pseudospin transfer torque acts.  The advantage of this 
approach is that experimental data could be used to obtain the spatial
distribution of pseudospin transfer torques.  Examination of coherent bilayer 
transport data suggests\cite{jpeahm} that the torques sometimes act along most
of the perimeter of the system, as assumed in a previous\cite{rossi:2004} attempt 
to understand coherent-bilayer tunneling data, and sometimes close to the source and drain
the contacts as assumed here.  The current model is most appropriate well away 
from the coherent state phase boundary as we have discussed.

The present version of the pseudospin transport torque theory does not account for 
thermal or quantum fluctuations of the condensate, which are unimportant in 
metallic ferromagnets but might sometimes be important in coherent bilayers.
Weak-coupling theories\cite{stern:2001,balents:2001,fogler:2001,brits} of bilayer tunneling
do account for fluctuations, but treat $\Delta_t$ perturbatively.  These theories 
cannot account for the existence of a critical currents and in practice assume that 
each layer has a separate well-defined local chemical potential.  It is clear from 
published transport data that this assumption is not always valid, in particular that 
it is not valid in the portion of the phase diagram far away from the phase boundary
on which the present paper focuses.  Experimentally\cite{Tiemann:2009} the critical current
value decreases as the phase boundary of the 
coherent state is approached by increasing either temperature or the effective 
layer separation $d/l$.  It would be interesting 
to attempt quantitative tests of the predictions of weak-coupling theories
in the portion of the phase diagram close to the phase boundary.   

The authors acknowledge essential contributions by T. Pereg-Barnea 
to initial stages of this work and insights gained from   
valuable discussions with A. Balatsky, J.P. Eisenstein, W. Dietsche, 
A.D.K. Finck, Y. Joglekar, D. Pesin, L. Radzihovsky, L. Tiemann, and K. von Klitzing.



\begin{thebibliography}{99}

\bibitem{fertig:1989} H. A. Fertig, Phys. Rev. B {\bf 40}, 1087 (1989). 

\bibitem{macdonald:1990a} A. H. MacDonald, P. M. Platzman, and G. S. Boebinger,
Phys. Rev. Lett. {\bf 65}, 775 (1990). 

\bibitem{wen:1992} X.-G. Wen and A. Zee, Phys. Rev. Lett. {\bf 69}, 1811 (1992).

\bibitem{eisenstein:2004} J. P. Eisenstein and A. H. MacDonald,
Nature {\bf 432}, 691 (2004). 

\bibitem{macdonald:1990b} A. H. MacDonald, E. H. Rezayi,
Phys. Rev. B {\bf 42}, 3224 (1990). 

\bibitem{macdonald:2002} A. H. MacDonald, Physica B {\bf 298}, 129 (2001). 
\bibitem{moon:1995} K. Moon, H. Mori, Kun Yang, S. M. Girvin, A. H. MacDonald, L. Zheng, D. Yoshioka, and Shou-Cheng Zhang, Phys. Rev. B {\bf 51}, 5138 (1995).

\bibitem{spielman:2000} I. B. Spielman, J. P. Eisenstein, L. N. Pfeiffer, and K. W. West, Phys. Rev. Lett. {\bf 84}, 5808 (2000).

\bibitem{jpeexpts}  I. B. Spielman, M. Kellogg, J. P. Eisenstein, L. N. Pfeiffer, and K. W. West,
Phys. Rev. B {\bf 70}, 081303(R) (2004);  A. D. K. Finck, A. R. Champagne, J. P. Eisenstein, L. N. Pfeiffer, and K. W. West, Phys. Rev. B {\bf 78},
075302 (2008). 

\bibitem{princeton}  E. Tutuc, S. Melinte, E. P. De Poortere, R. Pillarisetty,  and M. Shayegan, Phys. Rev. Lett. {\bf 91}, 076802 (2003);
E. Tutuc, M. Shayegan and D. A. Huse, Phys. Rev. Lett.
{\bf 93}, 036802 (2004).

\bibitem{stuttgart} R. D. Wiersma, J. G. S. Lok, S. Kraus, W. Dietsche, K. von Klitzing, 
D. Schuh, M. Bichler, H. P. Tranitz, and W. Wegscheider,
Phys. Rev. Lett. {\bf 93}, 266805 (2004);
L. Tiemann, J. G. S. Lok, W. Dietsche, K. von Klitzing, K.
Muraki, D. Schuh, and W. Wegscheider, Phys. Rev. B {\bf 77}, 033306 (2008).



\bibitem{stern:2001} A. Stern, S.M. Girvin, A. H. MacDonald, and Ning Ma, Phys. Rev. Lett. {\bf 86}, 1829 (2001). 

\bibitem{balents:2001} L. Balents and L. Radzihovsky, Phys. Rev. Lett. {\bf 86}, 1825 (2001). 

\bibitem{fogler:2001} M. M. Fogler and F. Wilczek,
Phys. Rev. Lett. {\bf 86}, 1833 (2001). 

\bibitem{brits} R. L. Jack, D. K. K. Lee, and N. R. Cooper, Phys. Rev. Lett.
{\bf 93}, 126803 (2004), {\em ibid} Phys. Rev. B 71, 085310 (2005);
O. G. C. Ros and D. K. K. Lee, arXiv:0911.2647.

\bibitem{spielman:2001} I. B. Spielman, J. P. Eisenstein, L. N. Pfeiffer, and K. W. West, Phys. Rev. Lett. {\bf 87}, 036803 (2001).
Instead of splitting the zero-bias conductance peak into two finite-bias peaks, a parallel field gradually decreases 
the its height.  Small features appear in the tails of these peaks 
near the voltages at which the conductances peaks are expected in the weak-coupling theory.

\bibitem{wen:1993} X. G. Wen and A. Zee, Phys. Rev. B {\bf 47}, 2265 (1993).

\bibitem{rossi:2004} E. Rossi, A. S. Nunez, and A. H. MacDonald, Phys. Rev. Lett. {\bf 95}, 266804 (2005).

\bibitem{khomeriki:2006} R. Khomeriki, L. Tkeshelashvili, T. Buishvili, and Sh. Revishvili,
Eur. Phys. J. B {\bf 51}, 421 (2006).

\bibitem{fil:2007} D. V. Fil and S. I. Shevchenko, J. Low Temp Phys. {\bf 33}, 780 (2007).

\bibitem{su:2008} J.-J. Su, and A. H. MacDonald, Nature Physics {\bf 4}, 799 (2008).

\bibitem{Stiles} See D. C. Ralph, and M. D. Stiles, J. Magn. Mag. Mater. {\bf 320}, 1190 (2008) and work cited therein.

\bibitem{slonczewski} J. C. Slonczewski, J. Magn. Magn. Mater. {\bf 159}, L1 (1996).

\bibitem{burkov} A. A. Burkov and A. H. MacDonald, Phys. Rev. B {\bf 66}, 115320 (2002).  

\bibitem{joglekar:2001} Y. N. Joglekar and A. H. MacDonald, Phys. Rev. B {\bf 64}, 155315 (2001). 

\bibitem{joglekar:2001a} Y. N. Joglekar and A. H. MacDonald, Phys. Rev. Lett. {\bf 87}, 196802 (2001). 
The analysis of tunneling transport in this paper is incomplete because of the 
absence of a pseudospin transfer torque term in the condensate equations of motion. 

\bibitem{phaseboundary} A. R. Champagne, J. P. Eisenstein, L. N. Pfeiffer, and K. W. West,
Phys. Rev. Lett. {\bf 100}, 096801 (2008); 
P. Giudici, K. Muraki, N. Kumada, Y. Hirayama, and T. Fujisawa,
Phys. Rev. Lett. {\bf 100}, 106803 (2008); 
A. D. K. Finck, J. P. Eisenstein, L. N. Pfeiffer, K. W. West,
arXiv:0911.2461. 

\bibitem{currentcaveat} Strictly speaking the statements are true only if the potential terms in the 
Schrodinger equation are local, a property not satisfied by exchange potentials.  This limitation is 
irrelevant since our goal is to describe systems with pseudospin textures that are smooth on 
microscopic length scales. 

\bibitem{nunez} A. S. Nunez and A. H. MacDonald, Solid State Comm., {\bf 139}, 31 (2006).

\bibitem{duinestt} See for example P. M. Haney, R. A. Duine, A. S. Nunez, and A. H. MacDonald,
J. Magn. Mag. Mater. {\bf 320}, 1300 (2008).  

\bibitem{champagne:2008} A. R. Champagne, A. D. K. Finck, J. P. Eisenstein, L. N. Pfeiffer, and K. W. West, Phys. Rev. B {\bf 78}, 205310 (2008). 

\bibitem{Tiemann:2009} L. Tiemann, Y. Yoon, W. Dietsche, K. von Klitzing, and W. Wegscheider, Phys. Rev. B {\bf 80}, 165120 (2009).

\bibitem{spielman:thesis} I. B. Spielman, Ph.D. Thesis, California Institute of Technology (2004).

\bibitem{sternhalperin} A. Stern and B. I. Halperin, Phys. Rev. Lett. {\bf 88}, 106801 (2002).

\bibitem{radzihovsky} M. Abolfath, A. H. MacDonald, and L. Radzihovsky,
Phys. Rev. B {\bf 68}, 155318 (2003). 

\bibitem{pesin} D. Pesin and A. H. MacDonald, in preparation. 

\bibitem{meronrefs} S. Q. Murphy, J. P. Eisenstein, G. S. Boebinger, L. N. Pfeiffer,
and K. W. West, Phys. Rev. Lett. {\bf 72}, 728 (1994).

\bibitem{yoshioka} D. Yoshioka and A. H. MacDonald, Phys. Rev. B {\bf 53}, R16168 (1996).

\bibitem{jpeahm} J. P. Eisenstein and A. H. MacDonald, unpublished.  

\bibitem{eastham:2009} P. R. Eastham, N. R. Cooper and D. K. K. Lee, Phys. Rev. B {\bf 80}, 045302 (2009). 

\bibitem{fertig:2005} H. A. Fertig and G. Murthy, Phys. Rev. Lett. {\bf 95}, 156802 (2005).

\bibitem{roostaie:2008} B. Roostaei, K. J. Mullen, H. A. Fertig, and S. H. Simon, Phys.
Rev. Lett. {\bf 101}, 046804 (2008).    

\end{thebibliography}
\end{document}